\documentclass[sigconf]{acmart}
\usepackage{xcolor}
\usepackage{graphicx}
\usepackage{braket}
\usepackage{subcaption}
\usepackage{algorithm}
\usepackage{algorithmic}
\usepackage{float}
\usepackage{quantikz}
\usepackage{microtype}
\usepackage{pifont}
\usepackage{bbding}
\usepackage{tcolorbox}
\usepackage{listings}
\usepackage{listings}
\usepackage{textgreek}
\usepackage{xcolor} 
\usepackage{amsmath}

\usepackage{amssymb}

\renewcommand\footnotetextcopyrightpermission[1]{} 
\begin{document}
\title[]{Adaptive Job Scheduling in Quantum Clouds Using Reinforcement Learning}

\author{Waylon Luo}
\affiliation{%
  \institution{Kent State University}
  \city{Kent}
  \state{OH}
  \country{USA}
}

\author{Jiapeng	Zhao}
\affiliation{%
  \institution{Cisco}
  \city{San Jose, CA}
  \country{USA}}

\author{Tong Zhan}
\affiliation{%
  \institution{Meta}
  \city{Menlo Park, CA}
  \country{USA}}

\author{Qiang Guan}
\affiliation{%
  \institution{Kent State University}
  \city{Kent}
  \state{OH}
  \country{USA}}

\begin{abstract}

Present-day quantum systems face critical bottlenecks, including limited qubit counts, brief coherence intervals, and high susceptibility to errors—all of which obstruct the execution of large and complex circuits. The advancement of quantum algorithms has outpaced the capabilities of existing quantum hardware, making it difficult to scale computations effectively. Additionally, inconsistencies in hardware performance and pervasive quantum noise undermine system stability and computational accuracy. To optimize quantum workloads under these constraints, strategic approaches to task scheduling and resource coordination are essential. These methods must aim to accelerate processing, retain operational fidelity, and reduce the communication burden inherent to distributed setups. One of the persistent challenges in this domain is how to efficiently divide and execute large circuits across multiple quantum processors (QPUs), especially in error-prone environments. In response, we introduce a simulation-based tool that supports distributed scheduling and concurrent execution of quantum jobs on networked QPUs connected via real-time classical channels. The tool models circuit decomposition for workloads that surpass individual QPU limits, allowing for parallel execution through inter-processor communication. Using this simulation environment, we compare four distinct scheduling techniques—among them, a model informed by reinforcement learning. These strategies are evaluated across multiple metrics, including runtime efficiency, fidelity preservation, and communication costs. Our analysis underscores the trade-offs inherent in each approach and highlights how parallelized, noise-aware scheduling can meaningfully improve computational throughput in distributed quantum infrastructures.

\end{abstract}
\maketitle

\section{Introduction}
\label{sec:introduction}

Quantum computing has the potential to transform fields such as cryptography, materials science, and machine learning by addressing problems beyond the capabilities of classical systems~\cite{Preskill_2018}. The foundational concept of using quantum machines to simulate complex physical phenomena was first introduced by Richard Feynman~\cite{feynman2018simulating}. Advances in both quantum hardware and algorithm design have recently brought this vision closer to reality, broadening the range of applications to areas such as pharmaceutical research~\cite{ZINNER20211680, vakili2024quantum}, financial analytics~\cite{9592468, egger2020quantum}, optimization problems, and machine learning tasks~\cite{8301126}.

Nevertheless, quantum computing still faces significant limitations due to hardware challenges, including restricted qubit numbers, fragile coherence, and substantial error rates. Quantum noise---originating from processes such as decoherence, gate infidelity, and measurement inaccuracies~\cite{Preskill_2018, Gambetta2017}---remains a major hurdle to achieving scalable computations. To address these issues, major quantum cloud platforms, such as IBM Quantum and Quantinuum, provide access to real-time calibration metrics, including coherence lifetimes, gate operation errors, and readout fidelity~\cite{Kandala2019, McKay2019}. Incorporating this dynamic information into scheduling techniques helps limit error propagation and improve the fidelity of quantum executions.

A key technical bottleneck for large-scale quantum computation is the constrained connectivity among qubits within a single processor. One emerging solution is to link multiple quantum processing units (QPUs) through classical communication networks. Vazquez et al.~\cite{Vazquez2024} demonstrated the practical execution of quantum circuits distributed across two 127-qubit processors interconnected via real-time classical channels. Although this marks important progress toward building scalable quantum platforms, the problem of optimally distributing and scheduling large quantum circuits across multiple QPUs remains largely unresolved.

Effective scheduling plays a crucial role in distributing large quantum circuits across multiple QPUs, aiming to optimize execution time, resource allocation, and circuit fidelity. In this work, we investigate four distinct scheduling approaches: (1) \textit{speed-optimized scheduling}, which focuses on minimizing runtime by uniformly distributing jobs across available QPUs without considering hardware-specific differences; (2) \textit{error-aware scheduling}, which dynamically routes jobs to QPUs with superior calibration metrics, including lower gate and readout errors, using real-time hardware data; (3) \textit{balanced scheduling}, which partitions circuits and allocates them evenly among all processors to promote balanced resource usage; and (4) \textit{reinforcement learning-based scheduling}, which applies machine learning techniques from the Gymnasium~\cite{gymnasium2023} framework to learn optimal allocation strategies. By incorporating hardware heterogeneity into the decision-making process, error-aware scheduling improves circuit fidelity while retaining high execution efficiency.

The primary contributions of this study are summarized below.

\begin{itemize}
    \item \textbf{Flexible quantum job orchestration} --- The framework supports both built-in and user-defined scheduling policies, along with integration of noise models to optimize performance for specific applications.
    \item \textbf{Adaptive, error-informed scheduling} --- We design and implement a dynamic scheduling method that uses calibration data to enhance circuit fidelity during execution.  
    \item \textbf{Evaluation metrics} --- We establish key benchmarks, including execution duration, fidelity levels, and communication overhead, to systematically assess scheduling effectiveness.  
    \item \textbf{Trade-off analysis between speed and fidelity} --- Our study highlights the inherent balance between minimizing runtime and maintaining computational accuracy in fast-execution and error-aware strategies.  
    \item \textbf{Publicly available simulation platform} --- We introduce the first open-source framework that supports scheduling of quantum circuits exceeding the capacity of a single QPU~\cite{ICPP2025}.
\end{itemize}

The remainder of this paper is structured as follows. Section~\ref{sec:related-work} provides an overview of quantum computing and related prior work. Section~\ref{sec:architecture} details the architectural design of the simulation framework. Section~\ref{sec:problem-definition} defines the core problems addressed in this study. Section~\ref{sec:Allocation-Strategies} introduces the proposed noise-aware scheduling strategies. Section~\ref{sec:performance} describes the performance evaluation metrics. Section~\ref{sec:case-study} presents the experimental evaluation and results. Finally, Section~\ref{sec:conclusion} concludes the paper by summarizing the main contributions and insights.

\section{Related Work}
\label{sec:related-work}

In quantum computing, information is encoded in quantum bits (qubits), which can exist in a superposition of basis states $\ket{0}$ and $\ket{1}$, allowing them to represent both classical values 0 and 1 simultaneously~\cite{Mermin2007}. Furthermore, qubits exhibit entanglement—a phenomenon where the state of one qubit becomes intrinsically linked to that of another, regardless of the physical distance between them. These fundamental properties allow quantum computers to address specific computational problems more efficiently than classical systems~\cite{Preskill_2018}.

A quantum circuit, composed of quantum gates, is executed on a quantum processing unit (QPU) by mapping logical qubits to physical qubits. During circuit execution, a sequence of gate operations is applied. At the conclusion of the computation, qubit measurements are performed~\cite{Mermin2007}. Today, researchers in machine learning~\cite{8301126}, pharmacological discovery~\cite{ZINNER20211680}, and financial modeling~\cite{9592468} are actively executing quantum circuits on quantum computers to explore applications such as molecular simulation, portfolio optimization, and risk analysis.

\begin{figure}[h!]
    \centering
    \includegraphics[width=0.42\textwidth]{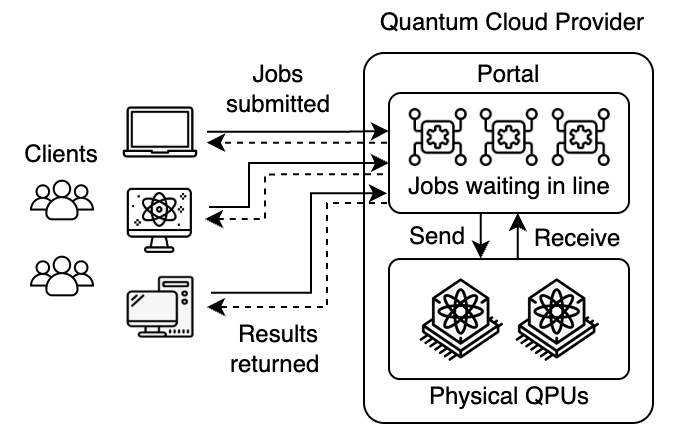}
    \caption{Overview of a quantum cloud computing system. }
    \label{fig:quantum_cloud}
\end{figure}

With advancements in quantum computing, Quantum Cloud Computing has emerged as a viable solution for remote access to quantum resources. It allows users to execute quantum algorithms on high-performance quantum processors via the cloud without owning costly hardware. Clients submit quantum jobs to a cloud-based portal, where jobs are queued and managed, as illustrated in Fig.~\ref{fig:quantum_cloud}. The portal dispatches jobs to available physical quantum processing units (QPUs) for execution. Upon completion, results are returned to the respective clients. This architecture captures the typical workflow and interaction between users and a quantum cloud provider. Companies such as IBM, Amazon Braket, and Microsoft provide public quantum platforms~\cite{PhysRevA.94.032329,Leymann2020QuantumIT,ravi2022quantum}, though these systems remain in the early stages compared to established cloud services like AWS and Azure. Simulations are critical to bridging this gap, allowing researchers to investigate quantum workloads, optimize resource allocation, and assess noise impacts in a cost-effective, flexible, and scalable manner.

Today’s QPUs have a limited number of qubits, which constrains the size of quantum circuits that can be executed natively. One strategy to address these limitations is \textit{circuit cutting}, where a large quantum circuit is partitioned into smaller subcircuits that run independently on smaller QPUs. Classical post-processing is then used to reconstruct the final result~\cite{Tang2021}. While circuit cutting is effective for certain problems, it introduces additional computational overhead and may be impractical when synchronous execution across multiple QPUs is required. In such cases, alternative techniques, such as real-time classical communication between QPUs, must be considered.

\begin{figure}[h!]
    \centering
    \includegraphics[width=0.48\textwidth]{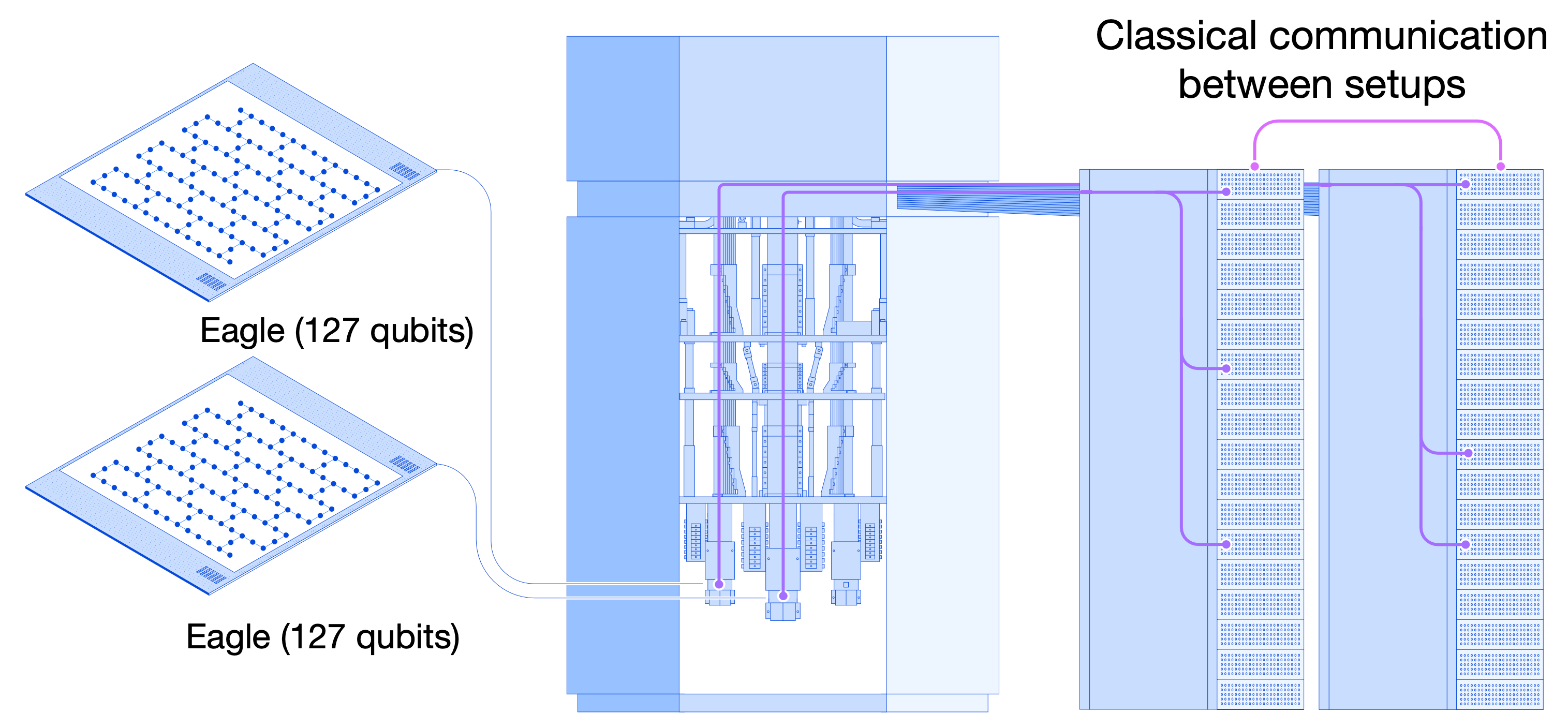}
    \caption{Connecting two QPUs via real-time classical communication. Image source: Vazquez et al. (2024) ~\cite{Vazquez2024}.}
    \label{fig:classical_communications}
\end{figure}
\
To address this limitation, Vazquez et al. conducted an experiment that combined two 127-qubit QPUs into a single virtual QPU using real-time classical communications~\cite{Vazquez2024}, as illustrated in Fig.~\ref{fig:classical_communications}. For example, executing a 150-qubit circuit on 127-qubit QPUs such as \textit{ibm\_eagle} requires splitting the circuit into smaller subcircuits and distributing them across two or more QPUs through classical communication channels. In this context, classical communication refers to the absence of quantum links between devices, relying instead on classical data transfer. This approach allows multiple processors to cooperate on a single computation by exchanging classical data mid-circuit, thereby extending the computational reach of current quantum hardware.

\begin{table*}[ht]
\centering
\caption{Comparison of Quantum Cloud and Quantum Simulation Frameworks}
\label{tab:framework_comparison}
\begin{tabular}{|l|l|l|c|c|}
\hline
\textbf{Framework} & \textbf{Research Focus} & \textbf{Research Methodology} & \textbf{Noise} & \textbf{Combined}\\
 & & & \textbf{Aware} & \textbf{QPUs} \\
\hline

QuNetSim~\cite{diadamo2021qunetsim} & Quantum network simulation & Discrete-event simulation & \ding{56} & \ding{56} \\
NetSquid~\cite{NetSquidAN} & Quantum networking and communication & Discrete-event simulation & \ding{51} & \ding{56} \\
QuEST~\cite{Jones2019-nc} & High-performance quantum state simulation & State-vector simulation & \ding{56} & \ding{56} \\
PAS~\cite{bian2023pas} & Lightweight quantum circuit simulation & State-vector simulation & \ding{56} & \ding{56} \\
QXTools~\cite{Brennan2022QXToolsAJ} & Distributed quantum circuit simulation in Julia & Distributed simulation & \ding{56} & \ding{56} \\
ProjectQ~\cite{ProjectQ} & Quantum programming framework & Compiler and circuit simulation & \ding{56} & \ding{56} \\
iQuantum~\cite{nguyen2023} & Quantum cloud modeling and simulation & Discrete-event simulation & \ding{56} & \ding{56} \\
QSimPy~\cite{nguyen2024qsimpy} & Learning-centric simulation for quantum clouds & Discrete-event simulation & \ding{56} & \ding{56} \\
QuCloud~\cite{liu2021qucloud} & Qubit mapping for multi-programming & Algorithmic modeling & \ding{56} & \ding{56} \\
QuMC~\cite{niu2024qumc} & Hardware-aware multi-programming compiler & Compiler-based simulation & \ding{51} & \ding{56} \\
Ravi et al.~\cite{ravi2023adaptive} & Quantum cloud job scheduling framework & Simulation-based evaluation & \ding{56} & \ding{56} \\
QURE~\cite{suchara2023qure} & Resource estimation in quantum clouds & Estimation modeling & \ding{56} & \ding{56} \\
QuTiP~\cite{JOHANSSON20131234} & Open quantum system dynamics simulation & Monte Carlo simulation & \ding{51} & \ding{56} \\
\textbf{This work} & Large-scale circuit simulation & Discrete-event simulation & \ding{51} & \ding{51} \\
\hline
\end{tabular}
\end{table*}

In the quantum computing field, simulations span from hardware-level noise modeling to high-level scheduling algorithm testing. Several existing frameworks have addressed various aspects of cloud and quantum simulation but differ significantly from our work in both scope and functionality. DRAS-CQSim~\cite{fan2021drascqsim} and GymCloudSim~\cite{jawaddi2024gymcloudsim} focus on classical high-performance cluster scheduling and energy-driven cloud scaling, respectively, without addressing quantum-specific or multi-QPU challenges. QuNetSim~\cite{diadamo2021qunetsim} and NetSquid~\cite{NetSquidAN} provide discrete-event simulations for quantum networks, where NetSquid incorporates noise models; however, both concentrate solely on network communication rather than distributed computation. QuEST~\cite{Jones2019-nc}, PAS~\cite{bian2023pas}, and QXTools~\cite{Brennan2022QXToolsAJ} emphasize quantum circuit simulation with high performance or distributed execution but do not address resource variability or scheduling in a cloud setting. Similarly, ProjectQ~\cite{ProjectQ} offers a quantum programming framework with circuit simulation capabilities but lacks cloud and scheduling support. iQuantum~\cite{nguyen2023} and QSimPy~\cite{nguyen2024qsimpy} propose quantum cloud simulation environments but emphasize modeling- and learning-centric frameworks without incorporating noise awareness or explicit job distribution across multiple QPUs. QuCloud~\cite{liu2021qucloud} and QuMC~\cite{niu2024qumc} address multi-programming on single quantum processors by optimizing qubit mapping and compiler techniques. Although QuMC incorporates hardware noise awareness, it remains limited to single-device settings. Ravi et al.~\cite{ravi2023adaptive} present a quantum job scheduling framework, yet without integrating noise models or real-time communication constraints. QURE~\cite{suchara2023qure} focuses on estimating resource requirements in quantum cloud systems rather than executing or scheduling jobs. QuTiP~\cite{JOHANSSON20131234} simulates open quantum system dynamics using Monte Carlo methods, providing noise modeling but targeting physical dynamics instead of distributed quantum computing. The related work comparison is presented in Table~\ref{tab:framework_comparison}.

These prior approaches either address quantum circuit execution, quantum networking, or classical workload management. Our simulation framework, based on the work by Vazquez et al.~\cite{Vazquez2024}, integrates real-time classical communication between distributed quantum processors. Unlike existing quantum circuit simulators, which assume monolithic QPU architectures, our framework models scenarios where quantum circuits exceed the capacity of a single QPU and must be partitioned and executed synchronously across multiple QPUs. This introduces additional challenges related to job scheduling, synchronization delays, and inter-processor communication—issues absent from current simulation tools. By addressing these gaps, our noise-aware simulation framework is the first to schedule quantum circuits that surpass the qubit capacity of a single device, providing a realistic representation of near-future cloud-based quantum computing.

\section{Simulation Architecture}
\label{sec:architecture}

\begin{figure}[h]   
    \includegraphics[width = 0.47\textwidth]{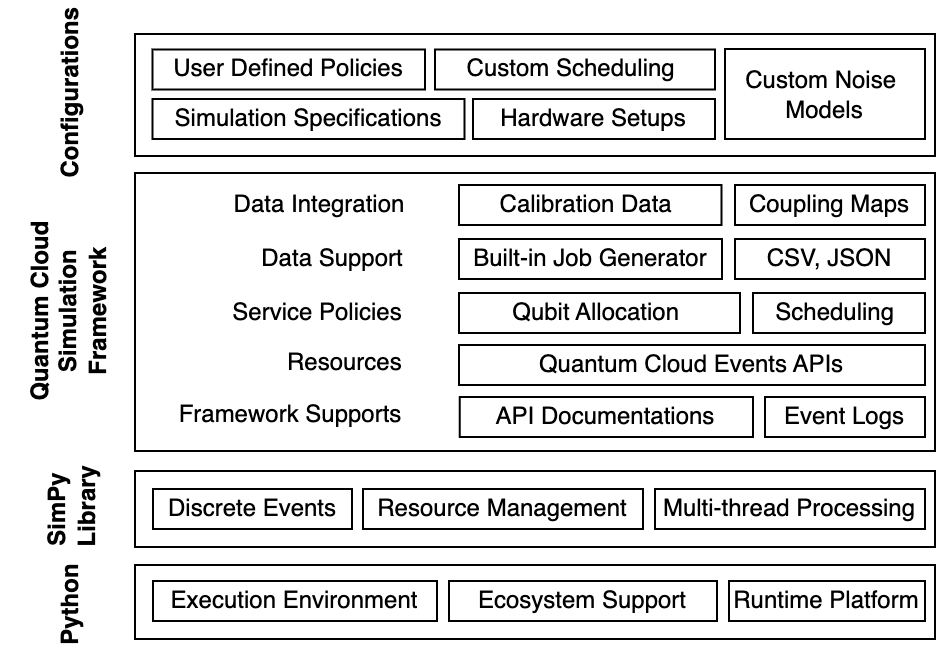}
    \caption{The architecture of the simulation framework consists of four layers: Python Lyer, SimPy Library, Quantum Cloud Simulation Framework, and Configurations.}
    \label{Software-Architecture}
\end{figure}

\begin{figure*}[h]
    \centering
    \includegraphics[width=\textwidth]{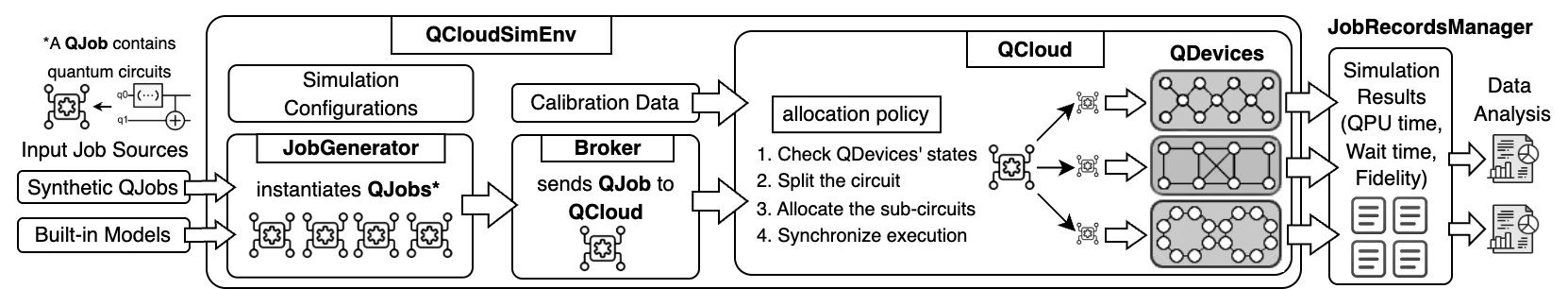} 
    \caption{The ecosystem of the framework simulates the end-to-end orchestration of quantum jobs, beginning from job sources (e.g., CSV/JSON files, or built-in models), through the framework environment—comprising the JobGenerator, Broker, and local queue management—to execution on quantum devices (QDevices). Simulation results are processed and visualized by the JobRecordsManager, providing insights into system performance and execution metrics.}
    \label{fig:ecosystem}
\end{figure*}

Our simulation framework adopts a layered architecture to model quantum cloud infrastructures. As illustrated in Fig.~\ref{Software-Architecture}, the framework comprises four main layers: the \textbf{Python Layer}, the \textbf{SimPy Library}, which provides the underlying simulation capabilities; the \textbf{Quantum Cloud Simulation Layer}, which models quantum cloud-specific components and operations; and the \textbf{Configurations Layer}, where users define simulation specifications.

The \textbf{Python Layer} underpins the framework by supporting the simulation ecosystem and providing a programming environment integrated with essential scientific libraries, including \texttt{NumPy} for numerical operations, \texttt{networkx} for graph organization, and \texttt{Matplotlib} for visualization and analysis.

The \textbf{SimPy Library}~\cite{SimPy} serves as the core event-driven simulation engine, managing discrete-event scheduling and resource allocation. Built on SimPy, our framework extends its functionality to support quantum cloud workload modeling and execution.

The \textbf{Quantum Cloud Simulation Layer} forms the backbone of the framework, modeling key quantum cloud components. This layer provides APIs for simulating core entities, framework utilities, and documentation to guide users in developing and extending the simulation. Customizable job scheduling and allocation strategies reside here. Users can implement tailored strategies based on optimization objectives. The layer also supports stochastic job generation and deterministic job flow through external data formats such as CSV and JSON. Centralized data management tracks job lifecycles, supporting post-simulation workload analysis. Furthermore, this layer integrates a library of QPU profiles, including coupling maps, device performance metrics, and calibration data.

The \textbf{Configurations Layer} sits at the top and provides interfaces for defining policies and simulation parameters without altering the core architecture. Users must specify scheduling and allocation policies, simulation parameters, and hardware configurations before running simulations.

The quantum cloud simulation environment orchestrates job flow through modular subcomponents.

The quantum cloud simulation environment, or \textbf{\texttt{QCloudSimEnv}}, extended from SimPy's \texttt{Environment}, serves as the core component of the framework. \texttt{QCloudSimEnv} consists of a \texttt{QCloud}, one or more \texttt{QDevice} instances, a \texttt{JobGenerator}, a \texttt{JobRecordsManager}, and a \texttt{Broker}.

The quantum cloud, or \textbf{\texttt{QCloud}}, initialized within \texttt{QCloudSimEnv}, manages quantum devices, allocates large circuits, and handles device communication. Allocation policies for large-scale circuits are implemented in this entity. These policies govern circuit partitioning and distribution across devices. In this work, jobs with large quantum circuits are partitioned into smaller circuits and allocated across \texttt{QDevices}.

The \textbf{\texttt{QDevice}} class, representing quantum devices and extended from \texttt{BaseQDevice}, defines qubit topology, operational characteristics, and resource management. The \texttt{QuantumDevice} subclass models graph-based qubit topologies to represent connectivity or coupling maps for superconducting devices. A specialized subclass, \texttt{IBM\_QuantumDevice}, further refines the model to capture IBM quantum hardware attributes, including Circuit Layer Operations Per Second \texttt{(CLOPS)}~\cite{IBMPerformance} and error scores derived from calibration data~\cite{IBMCalidocument}.

The \textbf{\texttt{Broker}} acts as the intermediary between job requests and available quantum devices, managing device selection, resource allocation, and job execution strategies. Users may create a \texttt{CustomBroker} by extending the abstract \texttt{Broker} class to implement custom algorithms and optimize scheduling according to specific objectives.

A quantum job, or \textbf{\texttt{QJob}}, encapsulates attributes and behaviors representing a quantum task. Each \texttt{QJob} instance includes several parameters: \texttt{job\_id}, a unique identifier; \texttt{num\_qubits}, the maximum number of qubits required; \texttt{depth}, the circuit depth; \texttt{num\_shots}, the number of repetitions per measurement; and \texttt{arrival\_time}, the job’s arrival time. \texttt{QJob} serves as an abstraction for simulating job scheduling and resource allocation on quantum devices. In this work, each \texttt{QJob} is assumed to contain a single quantum circuit.

The \textbf{\texttt{JobGenerator}} component produces \texttt{QJobs} through predefined dispatching mechanisms. For deterministic simulations, jobs can be loaded from external CSV files that specify fields such as job ID, number of shots, arrival time, circuit depth, and qubit requirements. The \texttt{JobGenerator} reads these data and schedules \texttt{QJobs} to arrive at the designated times. If no arrival time is specified, the current timestamp is assigned by default. This deterministic mode supports benchmarking, debugging, and comparative performance analysis under controlled conditions.

The \textbf{\texttt{JobRecordsManager}} component tracks the lifecycle of quantum jobs, logs key events, and maintains records of system activity. It monitors job-related events, including \texttt{arrival}, \texttt{start}, \texttt{finish}, and \texttt{fidelity}, which provide data for analyzing performance metrics such as wait times, execution durations, and throughput.

\textbf{Calibration data} provides real-time information about the performance of a quantum processor, including qubit coherence times, gate fidelities, readout errors, and other hardware metrics. These parameters are critical for assessing the reliability of quantum computations.

Together, these components define the framework's ecosystem. As shown in Fig.~\ref{fig:ecosystem}, the framework integrates multiple modules to manage the generation, allocation, execution, and analysis of quantum jobs (\texttt{QJob}s). Jobs can originate from various sources, including standardized benchmarks, synthetic datasets, or user-defined models. These diverse sources provide flexibility for a range of research scenarios. The central component, \texttt{QCloudSimEnv}, coordinates job flow through its modular subcomponents. The \texttt{JobGenerator} creates \texttt{QJob}s with defined circuits and metadata, including resource requirements and execution parameters. These jobs are passed to the \texttt{Broker}, which schedules them based on customizable allocation policies and dispatches them to quantum devices within the \texttt{QCloud}. Each \texttt{QDevice} is configured with specific characteristics, including qubit connectivity, gate fidelity, and noise properties, and maintains a local queue for job execution. Allocation policies determine how jobs are distributed across devices to optimize system performance. The \texttt{JobRecordsManager} collects simulation results, capturing execution metrics such as QPU time, wait time, and fidelity. These records support post-simulation analysis, helping researchers evaluate and refine quantum cloud resource management strategies.
\section{Problem Definition}
\label{sec:problem-definition}
Efficient allocation of large quantum jobs across multiple quantum processors is a central challenge in distributed quantum computing. This problem is shaped by limited qubit capacity, hardware variability, and fidelity loss incurred during inter-device communication. We formally define the allocation problem and present a reinforcement learning-based approach to address it. Firstly, we consider a quantum cloud comprised of a set of heterogeneous quantum processing units (QPUs), denoted
\[
  \mathcal{D} = \{D_1, D_2, \ldots, D_{|\mathcal{D}|}\}.
\]
Each device \(D_i\in\mathcal{D}\) is characterized by the tuple
\[
  D_i = \bigl(C_i,\,E_i,\,K_i,\,G_i\bigr),
\]
where:
\begin{itemize}
  \item \(C_i\in\mathbb{N}\) is the qubit capacity (i.e., the number of available qubits, represented in our code by \texttt{device.container.level}).
  \item \(E_i\in[0,1]\) is the device’s error score (e.g., a combination of single‑ and two‑qubit error rates).
  \item \(K_i\in\mathbb{R}^+\) is the device throughput in circuit layer operations per second (CLOPS)~\cite{IBMPerformance}.
  \item \(G_i=(V_i, E_i')\) is the qubit connectivity graph, with \(|V_i|=C_i\).
\end{itemize}

A \emph{job} \(J\) arriving to the cloud is defined by
\[
  J = \bigl(q,\,d,\,s,\,t_2\bigr),
\]
where:
\begin{itemize}
  \item \(q\in\mathbb{N}\) is the total number of qubits required.
  \item \(d\in\mathbb{N}\) is the circuit depth.
  \item \(s\in\mathbb{N}\) is the number of shots to execute.
  \item \(t_2\in\mathbb{N}\) is the number of two‑qubit gates.
\end{itemize}

An \emph{allocation} of \(J\) to a subset of \(k\) devices is a vector
\[
  \mathbf{a} = \bigl(a_1, a_2, \ldots, a_k\bigr)
  \quad\text{such that}\quad
  \sum_{i=1}^k a_i = q
  \quad\text{and}\quad
  a_i \le C_{\,\pi(i)},
\]
where \(\{\pi(1),\dots,\pi(k)\}\subseteq\{1,\dots,|\mathcal{D}|\}\) indexes the selected devices.  
We further require that each device \(D_{\pi(i)}\) contains a connected subgraph of size \(a_i\), i.e.,
\[
  \exists\,S_i\subseteq V_{\pi(i)},\;|S_i|=a_i,\;
  G_{\pi(i)}[S_i]\text{ is connected.}
\]

Once allocated, the job incurs:
\begin{itemize}
  \item \textbf{Processing time} on device \(D_{\pi(i)}\):
  \[
    T_i = \frac{M\,K\,s\,\log_2(QV_{\pi(i)})}{K_{\pi(i)}\,\times\,60},
  \]
  where \(M\) and \(K\) are constants capturing IBM‑specific metrics (cf.\ \(\tt calculate\_process\_time\)), and \(QV_{\pi(i)}\) is the quantum volume of the device \(D_{\pi(i)}\).
  The overall job runtime is
  \(\displaystyle T(\mathbf{a}) = \max_{i=1,\dots,k} T_i.\)

  \item \textbf{Device fidelity} on \(D_{\pi(i)}\):
  \[
    F_i = \bigl(1 - \epsilon_{\mathrm{1q}}\bigr)^d
        \;\times\;\bigl(1 - \epsilon_{\mathrm{ro}}\bigr)^{\sqrt{a_i}}
        \;\times\;\bigl(1 - \epsilon_{\mathrm{2q}}\bigr)^{\sqrt[4]{t_2}},
  \]
  where \(\epsilon_{\mathrm{1q}},\epsilon_{\mathrm{ro}},\epsilon_{\mathrm{2q}}\) are the device’s average single‑qubit, readout, and two‑qubit errors respectively.
\end{itemize}

Devices must then exchange classical and quantum data; each of the \(k-1\) inter‑device links imposes a communication penalty
\[
  P = \beta^{\,k-1},
  \quad
  \beta\in(0,1),
\]
so that the \emph{final job fidelity} is
\[
  F(\mathbf{a}) = \Bigl(\tfrac{1}{k}\sum_{i=1}^k F_i\Bigr)\;\times\;P.
\]

Given a job \(J\) and the device set \(\mathcal{D}\), our goal is to find an allocation \(\mathbf{a}\) and a choice of devices \(\{\pi(1),\ldots,\pi(k)\}\) that simultaneously
\[
  \max_{\mathbf{a},\,\pi(\cdot)}\;F(\mathbf{a})
  \quad\text{and}\quad
  \min_{\mathbf{a},\,\pi(\cdot)}\;T(\mathbf{a}),
\]
subject to the capacity and connectivity constraints above.

Since this work focuses on large-scale quantum circuits, the jobs are prepared with certain size of circuits. The constraint for required qubits for a job /textit{i} is expressed as:
\
\begin{equation}
\max(|C|_1, |C|_2, \dots, |C|_N) < q_i < \sum_{j=1}^{N} |C|_j.
\end{equation}

In that way, we make sure that all circuits are large enough to require partitioning and allocating across multiple quantum devices while remaining small enough to fit within the total available qubits in the quantum cloud $\mathbb{Q}$.

\subsection{Reinforcement Learning Formulation}

To address the complexity of allocating large quantum jobs across heterogeneous quantum devices, we formulate the problem as a Markov Decision Process (MDP) and implement a reinforcement learning (RL) agent using Proximal Policy Optimization (PPO). The environment is modeled using the Gymnasium API with the following components:

\textbf{State:} A continuous vector consisting of normalized job and device features. Specifically:
    \begin{itemize}
    \item Job parameters: normalized qubit count \( q/q_{\max} \), where \( q_{\max} = 50 \).
    \item Device features for each of the \( k = 5 \) devices (padded with zeros if fewer): normalized container level \( C_i / 150 \), error score \( E_i \), and normalized CLOPS \( K_i / 10^6 \).
    \end{itemize}
The resulting state vector has dimensionality \( 1 + 3k = 16 \).

\textbf{Action:} A continuous vector \( \mathbf{a} = [a_1, \dots, a_k] \) representing unnormalized allocation weights. The final allocation is computed as:
\[
\hat{a}_i = \frac{a_i}{\sum_j a_j + \varepsilon} \cdot q
\]
followed by rounding and adjustment to ensure \( \sum_i \hat{a}_i = q \).

\textbf{Reward:} The scalar reward is the average circuit fidelity achieved across the allocated devices:
\[
R = \frac{1}{k'} \sum_{i=1}^{k'} F_i
\]
where \( k' \) is the number of devices used, and \( F_i \) is the fidelity on device \( i \), incorporating gate error, readout error, and (optionally suppressed) two-qubit error.

The RL agent is trained in simulation by generating randomized jobs and interacting with the environment over single-step episodes. Once trained, the policy network is deployed to produce allocation decisions in simulation: given a new job and the current system state, the model outputs allocation ratios, which are then used to request and reserve resources across multiple QPUs. After quantum job execution, inter-device communication is simulated (if applicable), and final fidelity is computed with a communication penalty. This approach supports adaptive, noise-aware allocation in distributed quantum systems.

\section{Allocation Strategies}
\label{sec:Allocation-Strategies}

Our framework supports four allocation modes that share a unified scheduling workflow. For each incoming quantum job, the scheduler selects a subset of devices, partitions the job accordingly, and orchestrates parallel execution with inter-device communication when necessary. The primary distinction among the allocation modes lies in the \textit{device selection policy}, which determines which devices are chosen for execution. 

\subsection{Allocation Workflow}

Algorithm~\ref{alg:unified_allocation} summarizes the unified allocation workflow shared across all scheduling modes. Once devices are selected using one of the policies, the subsequent steps — including qubit partitioning, parallel execution, inter-device communication, and fidelity computation — follow a common execution path.

\begin{algorithm}
\caption{Unified Allocation Workflow}
\label{alg:unified_allocation}
\begin{algorithmic}[1]

\STATE \textbf{Input:} Quantum job $J$, set of available devices $D$
\STATE \textbf{Output:} Allocation and execution of $J$ across selected devices

\STATE Select allocation mode (Speed-based, Error-aware, Fair, RL-based)
\STATE Identify candidate devices $D_{\text{selected}} \subseteq D$ based on allocation mode policy
\STATE Partition job $J$ into sub-jobs according to the number and capacity of $D_{\text{selected}}$

\FOR{each device $d$ in $D_{\text{selected}}$}
    \STATE Request and reserve necessary qubits from $d$
    \STATE Execute the assigned sub-job on $d$
\ENDFOR

\FOR{each pair of devices with dependent sub-jobs}
    \STATE Perform classical communication to synchronize results
\ENDFOR

\STATE Compute final fidelity, apply communication penalties if needed
\STATE Release qubits and log completion of $J$

\end{algorithmic}
\end{algorithm}

\textbf{Speed-based Mode.} This policy prioritizes minimizing total execution time by selecting devices with the fastest processing capability, without considering noise levels.

\textbf{Error-aware Mode.} This policy aims to maximize circuit fidelity by selecting devices with the lowest gate and readout errors based on calibration data. The calculation of the error score is detailed in Section~\ref{sec:error score}. Selected devices are sorted to prioritize those with lower error scores.

\textbf{Fair Mode.} This policy balances load by selecting devices with the lowest current utilization, aiming to prevent resource contention and evenly distribute workloads.

\textbf{Reinforcement Learning Mode.} This policy leverages a trained reinforcement learning agent to select devices based on real-time system state observations. The RL agent outputs allocation fractions that reflect learned trade-offs between device availability, error rates, and workload distribution.

\subsection{Qubit Partitioning and Allocation}

Once devices are selected, the job’s qubits are partitioned and assigned to available devices. Ideally, the subset of allocated qubits on each device forms a connected subgraph within the device's qubit topology graph \( G(V, E) \). However, finding an optimal connected subgraph is computationally intractable due to the combinatorial explosion of possible subgraphs. 

For example, selecting 10 qubits out of 127 requires searching over \( C(127, 10) = 209,123,798,385,425 \) combinations. To avoid prohibitive computational costs, our implementation adopts a simplified black-box abstraction that assumes allocated qubits form a connected subgraph, which is a practical assumption for devices with high connectivity. This approach enables fast and scalable allocation while focusing on evaluating broader scheduling performance.

\subsection{Complexity Analysis}
The computational complexity of the allocation workflow is primarily determined by the device selection and job partitioning steps. Device selection requires evaluating all available devices, resulting in a complexity of \(O(n)\), where \(n\) is the number of candidate devices. The job partitioning step divides the job across the selected devices, which operates in linear time with respect to the number of selected devices, yielding a complexity of \(O(m)\), where \(m\) is the number of selected devices. Execution and communication steps are simulated in an event-driven fashion and do not introduce significant overhead in the scheduling process. Therefore, the overall complexity of the algorithm is dominated by device evaluation and selection, and can be expressed as \(O(n + m)\). This makes the workflow efficient and scalable for quantum cloud environments, where the number of devices is typically moderate.

\subsection{Error Score}
\label{sec:error score}

Calibration data~\cite{IBMCalidocument} provides real-time information about the performance of a quantum processor, including qubit coherence times, gate fidelities, readout errors, and other hardware metrics. Based on this data, we define an \textit{error score} to quantify the overall device quality by combining readout errors, single-qubit gate errors, and two-qubit gate errors through a weighted formula:
\
\begin{equation}
\text{error\_score} = \alpha \cdot \frac{\sum_{i} \varepsilon_{\text{readout}, i}}{N_{\text{readout}}} + 
\theta \cdot \varepsilon_{\text{1Q}} +
\gamma \cdot \frac{\sum_{j} \varepsilon_{\text{2Q}, j}}{N_{\text{2Q}}}
\end{equation}
\
where:
\begin{itemize}
    \item $\varepsilon_{\text{readout}, i}$ represents the readout error for qubit $i$, and $N_{\text{readout}}$ is the number of qubits in the device.
    \item $\varepsilon_{\text{1Q}}$ is the error rate of the single-qubit RX gate.
    \item $\varepsilon_{\text{2Q}, j}$ denotes the two-qubit gate error for gate $j$, and $N_{\text{2Q}}$ is the total number of two-qubit gates.
\end{itemize}

The weighting factors $\alpha$, $\theta$, and $\gamma$ are assigned values of 0.5, 0.3, and 0.2, respectively. Readout errors are assigned the highest weight because they directly impact the correctness of measurement outcomes, making them especially critical to the fidelity of quantum computations. Single-qubit and two-qubit gate errors are weighted lower, with single-qubit errors receiving slightly higher weight. This reflects the observation that although two-qubit gates tend to have higher individual error rates, they may appear less frequently depending on circuit structure, while single-qubit gates are more common. This balance ensures that both the severity of errors and the frequency of operations are appropriately considered. The weighting scheme follows conventions in prior quantum error characterization studies~\cite{magesan2012characterizing} and can be adjusted as necessary for different quantum workloads.

\section{Performance Metrics}
\label{sec:performance}
In this work, we implemented and tested two allocation algorithms by evaluating their performance using three metrics: execution time, fidelity, and communication overhead.

\subsection{Execution Time}
\label{sec:execution}

The execution time (\(\tau\)) of a quantum job is computed based on the Circuit Layer Operations Per Second (CLOPS) and quantum volume (QV), metrics established by IBM~\cite{IBMPerformance}. CLOPS is a benchmark that measures the speed at which a quantum processor executes quantum circuits. QV evaluates the capability of a quantum system by considering factors such as qubit count, error rates, connectivity, and gate fidelity and determines the largest circuit depth a device can reliably handle. The execution time is given by:
\
\begin{equation}
    \tau = \frac{M \cdot K \cdot S \cdot D}{\textit{CLOPS}}
    \label{eq:tau}
\end{equation}
\
where:
\begin{itemize}
    \item \(M\): Number of circuit templates.
    \item \(K\): Number of parameter updates.
    \item \(S\): Number of shots for statistical accuracy.
    \item \(D\): Number of QV layers (\(\log_2\)QV).
\end{itemize}

The numerator (\(M \cdot K \cdot S \cdot D\)) in Equation \ref{eq:tau} represents the total computational workload, accounting for the number of circuits, qubits, iterations, and circuit depth. The denominator (\(\textit{CLOPS}\)) is a normalization factor, reflecting the quantum processor’s execution speed. For instance, consider a quantum job with the following parameters: \(M = 100\), \(K = 10\), \(S = 40{,}000\), and \(D = 7\) (circuit depth). The values for \(M\) and \(K\) are referenced from~\cite{IBMPerformance}. With \texttt{ibm\_brussels} QPU, which has a \(\text{CLOPS}\) rating of \(220{,}000\), the estimated execution time for this job applying Eq.~(\ref{eq:tau}) is approximately 21 minutes.

\subsection{Fidelity}
\label{sec:fidelity}
We calculate fidelity based on the calibration data provided by IBM. Error sources include single-qubit gate errors, two-qubit gate errors and readout errors. We estimate the overall fidelity using the following formulation:

\textbf{1. Single-Qubit Fidelity}: Calculated from single-qubit error which arises during the execution of single-qubit gates, such as Pauli-X or Hadamard gates, the single-qubit fidelity is estimated as:

\begin{equation}
F_{\text{1Q}} = (1 - \bar{\varepsilon}_{\text{1Q}})^{d}
\label{eq:1QFidelity}
\end{equation}

where:
\begin{itemize}
    \item $\bar{\varepsilon}_{\text{1Q}}$ is the average error rate of single-qubit gates.
    \item $d$ is the depth of the quantum circuit.
\end{itemize}

This follows the assumption that single-qubit gate errors compound independently over multiple gate applications, as described in quantum error models \cite{nielsen2010quantum}.

\textbf{2. Two-Qubit Fidelity}: The execution of two-qubit gates such as CNOT or CZ gates also arises errors. From those errors, the two-qubit fidelity is estimated as:

\begin{equation}
F_{\text{2Q}} = (1 - \bar{\varepsilon}_{2Q})^{\sqrt{N2Q}}, 
\label{eq:2QFidelity}
\end{equation}

where $\bar{\varepsilon}_{2Q}$ is the error rate for two-qubit gates that belong to the connectivity of two qubits in graph $G$ and $N2Q$ is the number of two qubits in the circuit. The Equation \ref{eq:2QFidelity}. computes the overall two-qubit fidelity as the product of individual gate survival probabilities, assuming independent errors for each two-qubit gate in the connectivity graph.

\textbf{3. Readout Fidelity}: Readout errors are caused by imperfect qubit measurement due to detector inefficiency or thermal condition of the device~\cite{ReadoutError}. Readout fidelity accounts for measurement errors and is computed as:
\
\begin{equation}
F_{\text{readout}} = (1 - \varepsilon_{\text{readout}})^{\sqrt{\frac{N_{\text{qubits}}}{N_{\text{devices}}}}}
\label{eq:ROFidelity}
\end{equation}

where:
\begin{itemize}
    \item $\varepsilon_{\text{readout}}$ is the average readout error per qubit.
    \item $N_{\text{qubits}}$ is the number of qubits in the circuit.
    \item $N_{\text{devices}}$ is the number of devices utilized.
\end{itemize}

This exponent scaling moderates the fidelity degradation compared to naive linear models, yielding more realistic estimates~\cite{magesan2012characterizing}.

\subsection{Overall Fidelity Estimation}

The fidelity of each quantum device, $F_{\text{dev}}$, is computed as the product of single-qubit, two-qubit, and readout fidelities, as shown in Equation~\eqref{eq:Fdev}:

\begin{equation}
F_{\text{dev}} = F_{\text{1Q}} \times F_{\text{2Q}} \times F_{\text{readout}}
\label{eq:Fdev}
\end{equation}

In multi-device scenarios, additional fidelity loss arises due to inter-device communication overhead.

\subsection{Inter-Device Communication Penalty}

Quantum jobs partitioned across multiple QPUs require classical communication of intermediate measurement outcomes or quantum state parameters. Due to technological limitations, our current model assumes classical communication of intermediate measurement outcomes between devices. To account for this, we introduce a communication penalty factor $\phi$:

\begin{equation}
F_{\text{final}} = \bar{F}_{\text{dev}} \times {\phi}^{(N{\text{devices}} - 1)}
\label{eq:F_final_penalty}
\end{equation}

Here, $\phi=0.95$ per inter-device connection follows empirical fidelity degradation observed experimentally in hybrid quantum computing setups~\cite{rigetti2018hybrid}. The variable $\bar{F}_{\text{dev}}$ denotes the average fidelity across associated quantum devices, and $N{\text{devices}}$ is the total number of interconnected quantum devices.

This simplified penalty-based model reflects typical fidelity degradation caused by classical communication latency, control synchronization issues, and inter-device calibration differences. Although simplified, this approach aligns with similar models employed in contemporary quantum computing research~\cite{Vazquez2024}, where fidelity degradation is empirically estimated.

\subsection{Communication Overhead Modeling}
\label{sec:communication_overhead}

Distributed quantum computing requires frequent classical communication, particularly when qubit measurements or classical control information must be transmitted between quantum processors during algorithm execution. Our model focuses specifically on classical communication latency, which is critical in current distributed quantum cloud architectures.

We model this classical communication overhead as proportional to the number of qubits involved and an experimentally motivated per-qubit latency parameter:

\begin{equation}
\tau_{\text{comm}} = N_{\text{qubits}} \cdot \lambda
\label{eq:comm_time}
\end{equation}

where:
\begin{itemize}
\item $N_{\text{qubits}}$ is the number of qubits whose measurement outcomes or classical control parameters are communicated between devices.
\item $\lambda$ is the per-qubit classical communication latency. The value is set to 0.02 seconds per qubit. It can vary based on existing networked quantum computer deployments. 
\end{itemize}

Equation~\eqref{eq:comm_time} adapts the classical communication complexity model~\cite{Yao1979}, explicitly focusing on classical data transfer between quantum devices. Communication is modeled as a blocking operation, delaying job execution by $\tau_{\text{comm}}$ before resuming computations.

This simplified latency-based model captures the essence of classical communication constraints prevalent in current experimental setups~\cite{rigetti2018hybrid, Vazquez2024}. However, it does not explicitly model more advanced quantum communication techniques such as entanglement swapping or teleportation, which remain experimental and challenging to implement at scale.

\subsection{RL Policy Training}
\label{sec:rl_training}

To implement the \textit{RL-based} allocation strategy, we trained a reinforcement learning policy using the Proximal Policy Optimization (PPO) algorithm. The training environment, \texttt{QCloudGymEnv}, was constructed to emulate real-world quantum cloud scheduling scenarios, with five IBM quantum processors (Strasbourg, Brussels, Kyiv, Québec, and Kawasaki) initialized using calibration data collected in March 2025. The objective of the reinforcement learning agent is to allocate quantum jobs to these devices so as to maximize the expected circuit fidelity while respecting device constraints.

\begin{figure}[ht]
    \centering
    \includegraphics[width=0.48\textwidth]{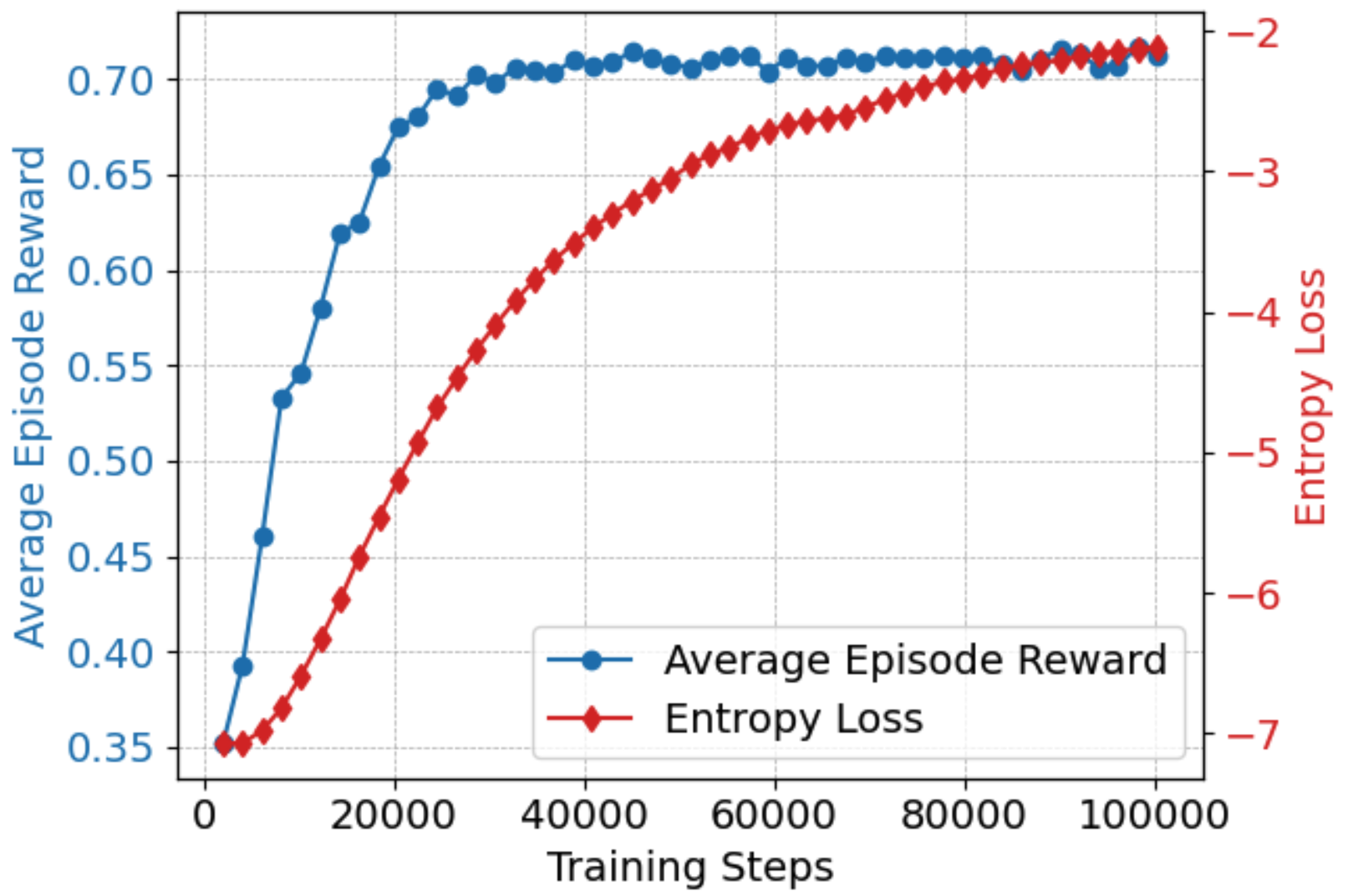}
    \caption{PPO training progress showing average episode reward (left y-axis) and entropy loss (right y-axis) over training steps.}
    \label{fig:ppo_training_progress}
\end{figure}

The training process was conducted over $100,000$ timesteps. Fig.~\ref{fig:ppo_training_progress} illustrates the training progress, showing the relationship between average episode reward and entropy loss. The average episode reward, which correlates with job fidelity, exhibits a steady improvement during the early stages of training and begins to plateau around $0.70$ as the model converges. Concurrently, entropy loss gradually decreases from approximately $-7$ to $-2$, indicating the agent's transition from exploration to exploitation as it learns more deterministic allocation policies.

The PPO agent was trained using a multi-layer perceptron (MLP) policy and default hyperparameters. The environment terminates each episode after a single allocation decision, making the problem a single-step decision-making task. As observed in the training curve, the learning process stabilized after around $40,000$ to $50,000$ timesteps, demonstrating that the agent was able to discover effective allocation strategies that yield consistently high fidelity across distributed devices.

Overall, the training results suggest that reinforcement learning is capable of automatically learning device allocation policies that balance fidelity and resource utilization without the need for hand-tuned heuristic rules. Further performance gains could potentially be achieved with extended training, curriculum learning, or communication-aware reward shaping, which are left as future work.

\section{Case Study}
\label{sec:case-study}

\begin{figure*}[ht]
  \centering
  \subfloat[Speed-Optimized]{\includegraphics[width=0.24\textwidth]{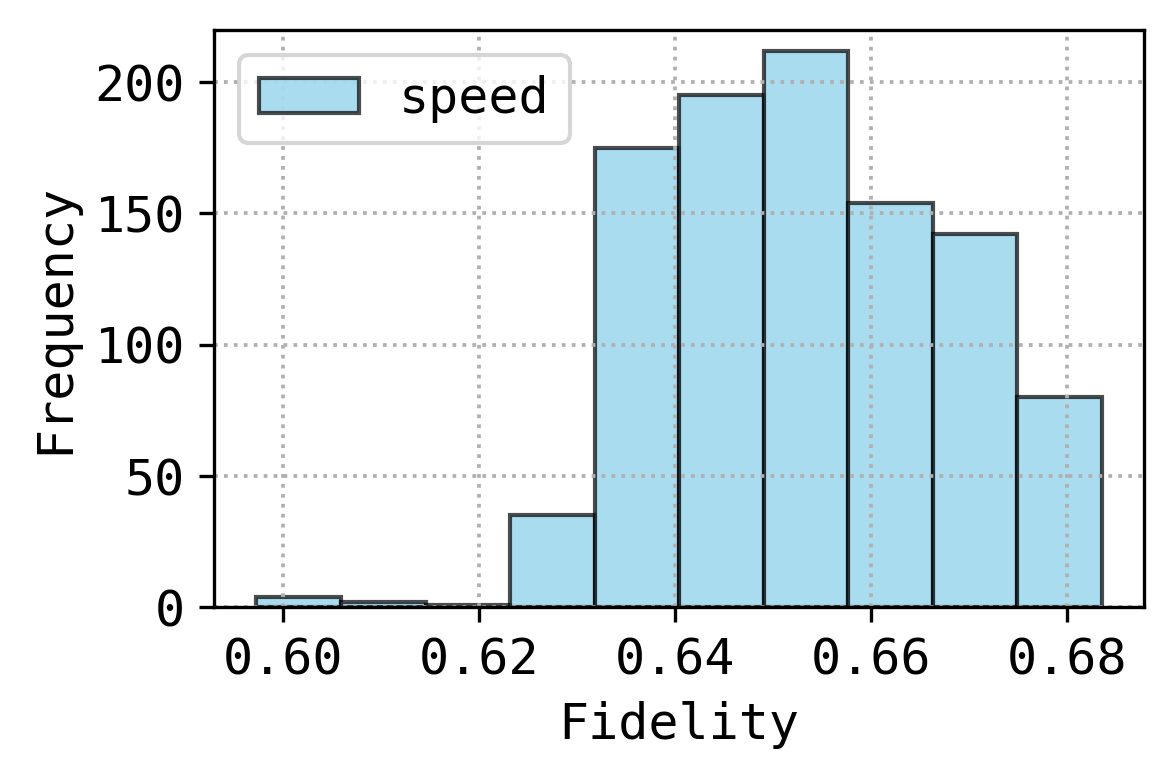}}
  \subfloat[Fidelity-Optimized]{\includegraphics[width=0.24\textwidth]{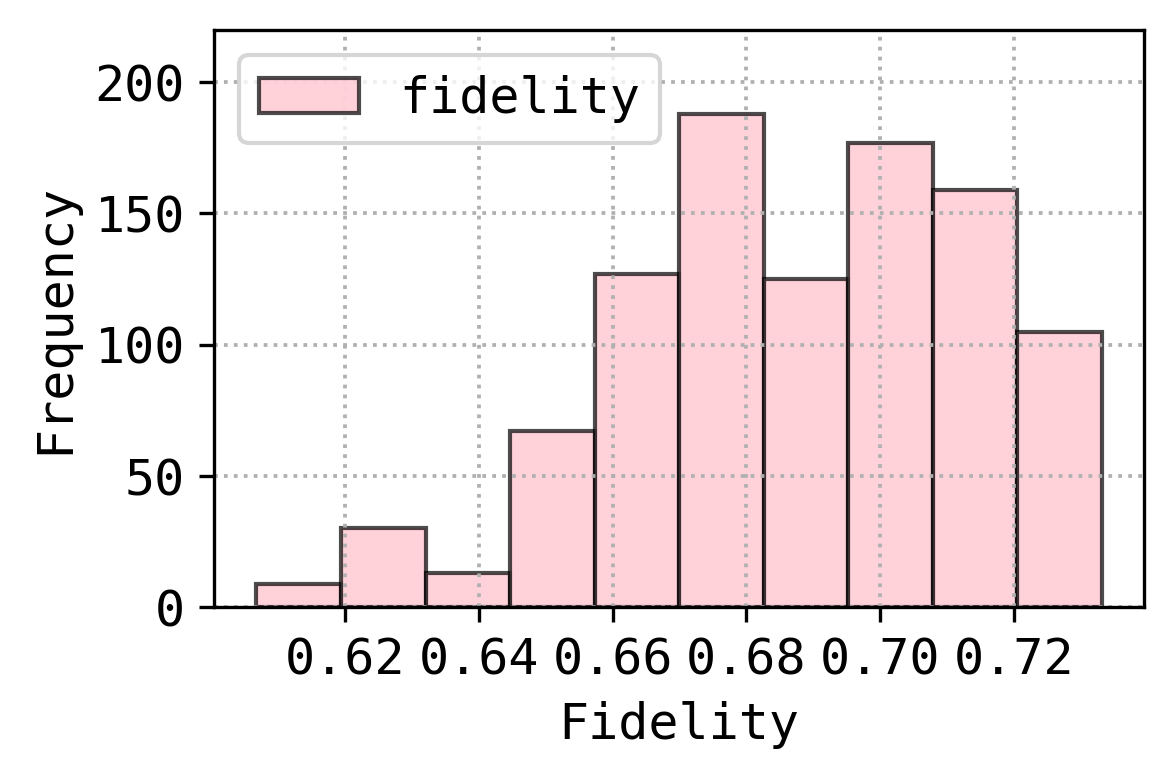}} 
  \subfloat[Fair Allocation]{\includegraphics[width=0.24\textwidth]{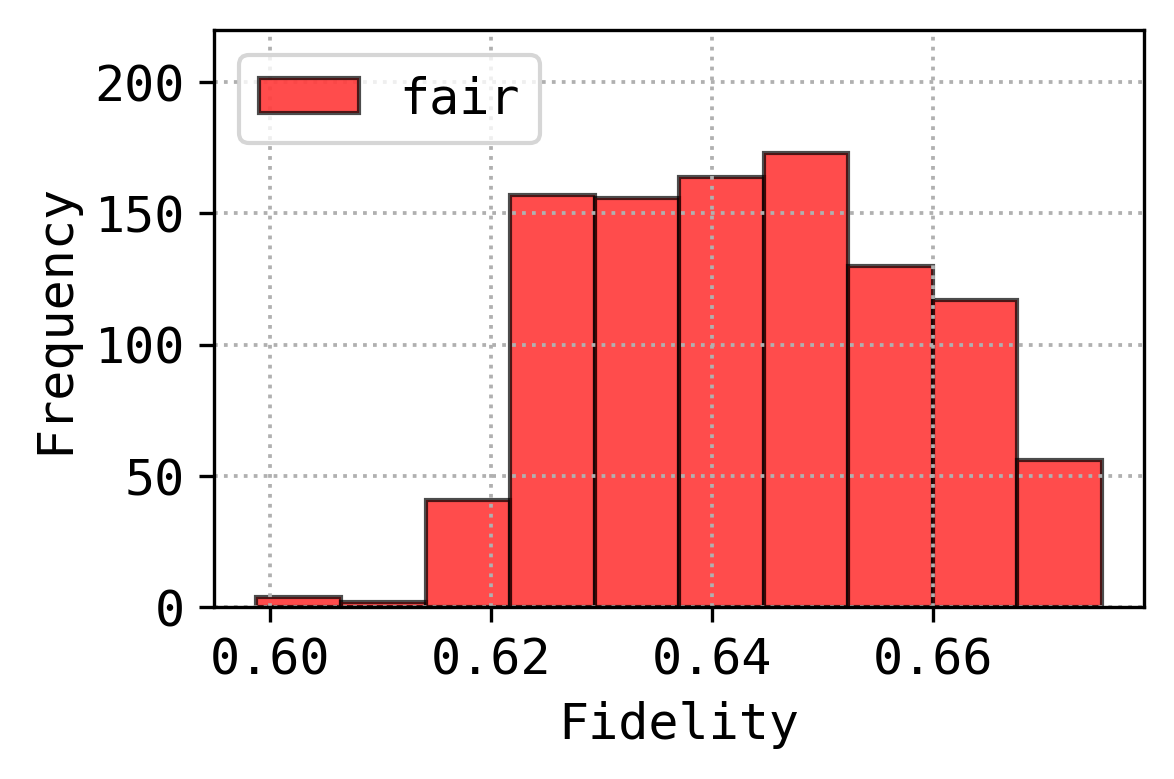}}
  \subfloat[RL-Based Allocation]{\includegraphics[width=0.25\textwidth]{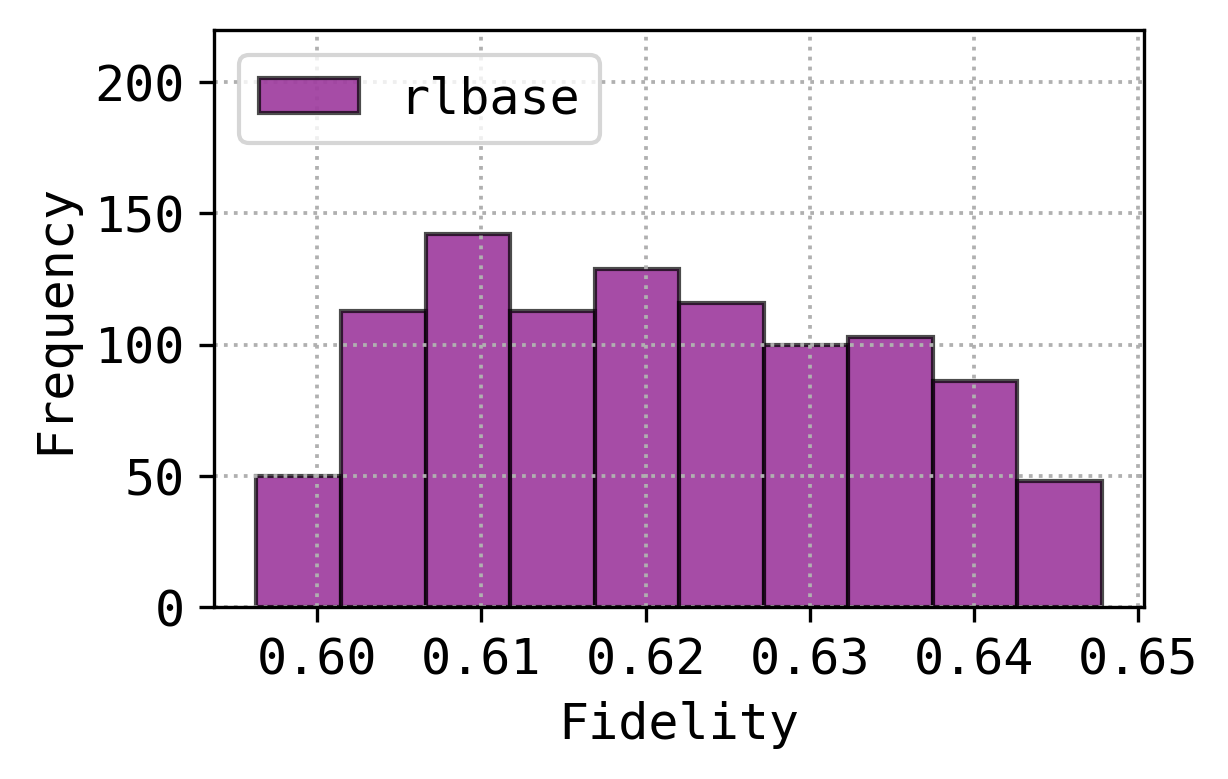}}

  \caption{Fidelity distributions of quantum jobs under four allocation strategies in the simulated quantum cloud environment. Each strategy demonstrates distinct fidelity patterns, reflecting trade-offs between speed, error-awareness, fairness, and reinforcement learning-based adaptive allocation.}
  \label{fig:fidelity_distributions}
\end{figure*}

In this section, we analyze the performance of four allocation modes. We evaluate performance metrics and identify the trade-offs among them. For this experiment, 1,000 synthetic jobs were generated. Each job requires between 130 and 250 qubits, has a circuit depth ranging from 5 to 20, and the number of shots between 10,000 and 100,000. The number of shots is chosen within a reasonable range to reflect typical execution settings in current quantum experiments. These synthetic jobs were specifically designed to necessitate splitting across multiple devices, thus emphasizing distributed execution. This method of intentionally creating jobs that exceed individual device capacities is relatively new, and as such, standard benchmarks for such large-scale, distributed quantum jobs have not yet been established in existing literature. The gate sets used in these jobs are abstracted to the number of single-qubit and two-qubit gates, without specifying explicit gate types. The jobs were tested using five simulated IBM quantum devices. All devices have 127 qubits and quantum volumes of 127. Among these devices, the QPUs \textit{ibm\_strasbourg} and \textit{ibm\_brussels} have the highest CLOPS (220,000), while \textit{ibm\_quebec}, \textit{ibm\_kawasaki}, and \textit{ibm\_kyiv} have CLOPS values of 32,000, 29,000, and 30,000, respectively~\cite{ibm_quantum_resources}.

\begin{table}[ht]
  \centering
  \begin{tabular}{llrrr}
    \toprule
    \textbf{Mode} & $T_{\mathrm{sim}}$ (s) & $\mu_F \pm \sigma_F$ & $T_{\mathrm{comm}}$ (s) \\
    \midrule
    speed    & 108\,775.38 & 0.65332 $\pm$ 0.01438 & 5\,707.80 \\
    fidelity & 209\,873.02 & 0.68781 $\pm$ 0.02605 & 3\,822.74 \\
    fair     & 108\,778.16 & 0.64373 $\pm$ 0.01478 & 5\,707.80 \\
    rlbase   & 106\,206.21 & 0.62087 $\pm$ 0.01301 & 6\,105.52 \\
    \bottomrule
  \end{tabular}
  \caption{Performance of allocation strategies on 1,000 large circuits.}
  \label{tab:allocation_results}
\end{table}

The performance of each strategy assessed in Table \ref{tab:allocation_results} is using the following metrics:
\begin{itemize}
  \item \textbf{Total simulation time} $T_{\mathrm{sim}}$: wall-clock time until all jobs complete (in seconds).
  \item \textbf{Average fidelity} $\mu_F \pm \sigma_F$: mean and standard deviation of final circuit fidelities.
  \item \textbf{Communication time} $T_{\mathrm{comm}}$: total inter-device communication delay summed over all jobs (in seconds).
\end{itemize}

\paragraph{Speed vs.\ Fidelity}
The \emph{speed} strategy minimizes simulation time ($1.09\times10^5$~s) by aggressively splitting jobs across all available QPUs, achieving moderate fidelity (0.6533) at the cost of high communication overhead (5.7~ks). In contrast, the \emph{fidelity} strategy prioritizes low-error devices, achieving the highest fidelity (0.6878), but with significantly longer runtime ($2.10\times10^5$~s). Its communication overhead is also lower (3.8~ks), indicating less fragmentation.

\paragraph{Fair Allocation}
The \emph{fair} strategy distributes workload evenly across devices, resulting in identical runtime to the \emph{speed} policy ($1.09\times10^5$~s) and slightly lower fidelity (0.6438). This suggests that equal partitioning may overlook hardware variability, underutilizing high-fidelity QPUs.

\paragraph{RL-Based Strategy}
The \emph{rlbase} policy, trained via PPO to maximize job fidelity, yields the shortest runtime (1.00$\times10^5$~s) but the lowest fidelity (0.6209) and highest communication overhead (6.3~ks). The flat fidelity distribution indicates that the learned strategy still explores allocations that cause excessive inter-device interaction, reducing fidelity despite faster job turnover.

\subsection{Fidelity Distributions}

The impact of each allocation strategy on the resulting job fidelity is illustrated in Fig.~\ref{fig:fidelity_distributions}. The \textit{Fair Allocation} and \textit{Speed-Optimized} strategies produce relatively narrow distributions, with fidelities concentrated around 0.65, indicating more deterministic but suboptimal fidelity outcomes. In contrast, the \textit{Fidelity-Optimized} strategy exhibits a right-shifted and bimodal distribution, successfully pushing a significant portion of jobs above 0.66 fidelity, which reflects its explicit focus on error-aware scheduling. Meanwhile, the \textit{RL-Based Allocation} strategy shows a flatter and broader distribution between 0.60 and 0.64. This pattern highlights the reinforcement learning agent’s tendency to explore diverse allocation configurations, trading off fidelity for potential adaptability. Overall, these fidelity profiles reveal the inherent trade-offs among speed, fairness, fidelity, and learning-driven allocation modes in distributed quantum job scheduling.

\subsection{Discussion}

The results highlight a fundamental trade-off in quantum-cloud scheduling between execution efficiency and output quality. The \emph{speed} and \emph{fair} strategies are effective in reducing overall simulation time, but this efficiency comes at the cost of lower fidelity in the resulting quantum circuits. In contrast, the \emph{fidelity}-based strategy prioritizes job quality by assigning workloads to devices with lower error rates, which improves circuit fidelity but leads to increased scheduling delays. The \emph{rlbase} model stands as a middle ground by balancing execution time and adaptive exploration. 

Overall, the observed fidelity improvements from the fidelity-aware strategy are modest. This limited difference likely arises from simplified fidelity estimation methods, which do not account for correlated noise or dynamic hardware variability, despite device heterogeneity and explicit inter-device communication penalties. All fidelity measurements in this work are based on theoretical estimations derived from reported error rates, qubit usage, and circuit depth. Experimental fidelity data from known quantum cloud services is not publicly available, as the distributed execution of quantum circuits that span multiple physical devices remains an emerging capability. 
\section{Conclusion}
\label{sec:conclusion}

In this work, we developed a simulation framework to explore quantum circuit execution across distributed quantum processors connected via real-time classical communication. Building on the architectural model proposed by Vazquez et al.~\cite{Vazquez2024}, our framework supports the coordinated scheduling of circuits that exceed the capacity of a single QPU, reflecting the anticipated shift toward multi-device execution in cloud-based quantum computing. We implemented and evaluated four distinct allocation strategies—\textit{Speed}, \textit{Fidelity}, \textit{Fair}, and \textit{Reinforcement Learning}—to examine trade-offs between execution time, circuit fidelity, and load distribution. The results demonstrate how different scheduling policies can significantly impact system performance and workload quality. As quantum hardware continues to advance, our framework provides a flexible foundation for simulating and analyzing scalable quantum job allocation strategies in distributed settings.

\bibliographystyle{unsrt}      
\bibliography{references}      
\end{document}